\documentclass[5p,numbered,sort&compress]{elsarticle}
\usepackage{graphicx}
\usepackage{hyperref}

\journal{Computer Physics Communications, 184 (12), 2703–2710 (2013)}
\begin{document}
\begin{frontmatter}

\title{{\tt MMonCa}: An Object Kinetic Monte Carlo simulator for damage irradiation evolution and defect diffusion}

\author[imdea]{Ignacio Martin-Bragado\corref{cor1}}
\ead{ignacio.martin@imdea.org}
\author[ifn]{Antonio Rivera}
\author[ifn]{Gonzalo Valles}
\author[imdea]{Jose Luis Gomez-Selles}
\author[dfa]{Mar\'ia J. Caturla}

\cortext[cpr1]{Corresponding author. Tel: +34 917871881. Fax: +34 915503047}
\address[imdea]{IMDEA Materials Institute, C/ Eric Kandel 2, 28906 Getafe, Madrid, Spain}
\address[ifn]{Instituto de Fusi\'on Nuclear, Universidad Polit\'ecnica de Madrid, Madrid, Spain}
\address[dfa]{Departamento de F\'isica Aplicada. Universitat d' Alacant. Alicante, Spain}

\begin{abstract} 
In this work we introduce the Object Kinetic Monte Carlo (OKMC) simulator {\tt MMonCa} and simulate the defect evolution in three different materials. We start by explaining the theory of OKMC and showing some details of how such theory is implemented by creating generic structures and algorithms in the objects that we want to simulate. Then we successfully reproduce simulated results for defect evolution in iron, silicon and tungsten using our simulator and compare with available experimental data and similar simulations. The comparisons validate {\tt MMonCa} showing that it is powerful and flexible enough to be customized and used to study the damage evolution of defects in a wide range of solid materials.

Work submitted to Computer Physics Communications, 184 (12), 2703–2710 (2013).

\url{http://dx.doi.org/10.1016/j.cpc.2013.07.011}
\end{abstract}


\end{frontmatter}

%\tableofcontents

\section{Introduction}

The study of irradiation effects and defect diffusion in solid materials is a field of the maximum importance given its implication in technological solutions for the microelectronic companies and as structural materials for nuclear fusion and fission energy generation. Many materials have been studied under irradiation. For metals there are studies on iron, tungsten \cite{RIETH-JNM11}, copper-niobium \cite{DEMKOWICZ-NIMB07,ZHANG-NIMB07,DEMKOWICZ-PRL08}, and others. For semiconductors silicon \cite{FAHEY-RMP89,PICHLER-BOOK04}, silicon carbide \cite{SNEAD-JNM96}, germanium \cite{BROWN-JAP59}, gallium arsenide \cite{ANHOLT-JAP88}, and others.

The physics involved within different crystalline solids when being irradiated is, to some degree, similar. Initially, irradiation produces a population of simple point defects, typically correlated in interstitial vacancy pairs called Frenkel pairs. After some initial recombination of these pairs, one of the constituents of the pair might diffuse faster at certain temperature ranges. For instance, interstitials for iron around 130\,K \cite{TAKAKI-RE83} or vacancies for silicon at room temperature \cite{WOODBURY-PRL60}. The moving particles agglomerate around clusters that, at some point, might evolve into extended defects with different shapes and properties \cite{COWERN-PRL99}. In some cases, the extended defects are dislocation loops \cite{BONINELLI-NIMB06}. Depending on the material, the extended defects might diffuse \cite{ARAKAWA-SCIENCE07} or be immobile \cite{CRISTIANO-JAP00}. When diffusing, it is possible for them to react with other defects or to reach the surface. When there are impurities present in the crystal, the impurities might diffuse in interstitial or substitutional positions, and form clusters by agglomerating with other impurities of the same species, with interstitials and vacancies, or even with different impurities. In some cases, the role of such impurities is crucial to understand the behavior of the material under irradiation. For instance, $He$ irradiation in metals, produces the formation of bubbles (He clusters) that are responsible for the change of the material mechanical properties in Fe \cite{SCHAUBLIN-JNM07}, Cu \cite{VELA-JNM66} W \cite{IWAKIRI-JNM02} and others. For semiconductor materials, clustering of dopants is responsible for electrical de-activation of species like As \cite{PANDEY-PRL88} and B \cite{PELAZ-APL97}.

The study of such important phenomena through computer simulations has been a field of research for decades. First principle calculations are used to obtain the activation energies and physical mechanisms of defect formation and diffusion \cite{FU-NATURE05,PANDEY-PRL88}. Lattice Kinetic Monte Carlo has been used to study macroscopic diffusivity \cite{CALISTE-PRL06}, cluster formation or recrystallization in heavily irradiated solids \cite{MARTIN-BRAGADO-APL09,LIU-MSEA97}. Object Kinetic Monte Carlo (OKMC) is one of the preferred tools used to study defect evolution inside solids \cite{JARAIZ-BOOK04,DOMAIN-JNM04,CATURLA-JNM00}. And finally, when the internal micro-structure of defects is assumed not to be important and the concentrations instead of the atomic positions offer enough information, finite element methods have been used. Many of these tools are well established and count with academic, open source, or commercial codes that are powerful and flexible enough to allow for fundamental research in all this wide range of materials. For instance, SIESTA \cite{SOLER-JPCM02}, VASP \cite{KRESSE-PRB93} or Gaussian \cite{GAUSSIAN-09} are available for first principle calculations. LAMMPS \cite{PLIMPTON-JCP95}, GROMACS \cite{BERENDSEN-CPC95} and many others are used to perform Molecular Dynamics, and there are many packages to run continuum (finite element method) simulations of which Abaqus and Ansys simulators, to name just a few, are well known and established. 

For OKMC, some existing codes are DADOS \cite{JARAIZ-BOOK04}, for diffusion of defects in silicon based materials, McDonalds \cite{JOHNSON-JAP98}, initially designed for silicon, Sentaurus Process KMC, a commercial software for Si based materials \cite{SENTAURUS-12}, and LAKIMOCA \cite{DOMAIN-JNM04}, a Lattice KMC used for simulation of irradiated metals. Nevertheless, there does not seem to be a clearly established, multi-material oriented, easy to access code for performing OKMC simulations. This lack could be negligible would it not have been for the extreme usefulness played by OKMC simulations in the field of damage irradiation: being in the border between atomistic and continuum simulation, OKMC plays a very important role in using all the theoretical information on activation energies obtained by the previously cited methods, and connecting them to macroscopic experiments \cite{FU-NATURE05,CASTRILLO-TNT01}. This is why in this work we want to introduce {\tt MMonCa}, a recent OKMC simulator written in C++ an integrated with the TCL \cite{TCL} script language, that wants to be multi-material, powerful, flexible and easy to use, filling the need for this type of codes that exist in the field of Monte Carlo simulation \cite{AMAR-CSE06}.

This article is structured as follows: We will start reviewing the KMC theory on Sec.~\ref{sec:theory} and the particular implementation of such theory on Sec.~\ref{sec:implementation}. Such implementation will review the major modules of {\tt MMonCa}: the time and space modules, (\ref{sec:time} and \ref{sec:space} respectively) and the description of all the implemented defect (object) types in Sec.~\ref{sec:defects}. The results and validations are written in Sec.~\ref{sec:results} starting with analytical calculations (\ref{sec:validation}) and then iron (\ref{sec:Fe}) silicon (\ref{sec:Si}) and tungsten (\ref{sec:W}). Finally, we will summarize the work in Sec.~\ref{sec:conclusions}.

\section{KMC theory}
\label{sec:theory}

The object Kinetic Monte Carlo algorithm goal is to follow the dynamic evolution of a system that might be out of equilibrium \cite{BORTZ-JCP75,VOTER-RES07,LANDAU-BOOK}. It assumes that there are different states in the system, and that the transitions between these states are Markovian, that is, that the transition rates $r_{ij}$ depend only on the initial $i$ state and the final $j$ state, and that such transitions are independent of time. These transitions $r_{ij}$ are the input parameters of the algorithm. In our particular case, we model them assuming Harmonic Transition State Theory \cite{VINEYARD-JPCS57} as Arrhenius laws with an activation barrier $E_{ij}$ (bigger than $k_BT$ for this approach to work) and a prefactor $P_{ij}$:
\begin{equation}
r_{ij} = P_{ij}\times\exp(-E_{ij}/k_BT).
\end{equation}

\begin{figure}
\begin{center}
\includegraphics{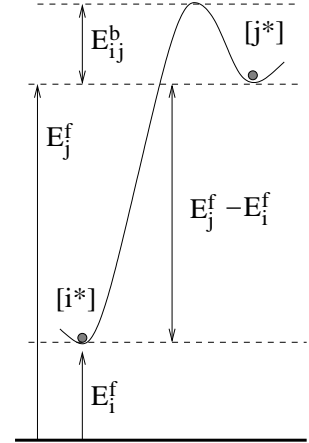}
\end{center}
\caption{Energetic diagram for our KMC simulations, showing two states $i$ and $j$ and the formation and barrier energies related with them.}
\label{fig:formations}
\end{figure}

The physical meaning of such barriers can be seen in Fig.~\ref{fig:formations}. In such diagram $E_{ij} = E_j^f - E_i^f + E^b_{ij}$. The opposite $E_{ji}$ would be just the energy $E_{ji} = E^b_{ij}$. Assuming that the concentration of particles in the $i$ state is $[i]$ and in the $j$ state is $[j]$, steady state will be reached when $[i] r_{ij} = [j] r_{ji}$. Using the notation stated in Fig.~\ref{fig:formations} and assuming $P_{ij} = P_{ji}$ we have that reaching such state implies 
\begin{equation}
[j^*]/[i^*] = \exp(-(E^f_j - E^f_i)/k_BT).
\label{eq:equilibrium}
\end{equation}

This relation does not include the barrier term $E^b_{ij}$, but the difference in formation energies. Out of steady state OKMC can take care of the dynamic behavior of the system and provide a way to simulate time evolution. The inclusion of interacting particles is not always exact and implies some assumptions and limitations. Such assumptions are, among others, a) a finite probability of trajectories to intersect without the reaction taking place for complex diffusion paths and/or complex object shapes, b) ternary reactions, c) collective movement and d) long-term reactions not happening or not being important. a) can be partly accounted for modifying the capture volumes, and b) and c) should not be a concern although there is a way to simulate collective movement in KMC~\cite{WHITELAM-JCP07}. Simulating long-term reactions is possible through the inclusion of quasi-continuum fields, for instance Fermi-level for Coulombic effects in semiconductors~\cite{MARTIN-BRAGADO-JAP05} and stress/strain computations for elastic interactions~\cite{ZOGRAPHOS-IIT12}. In this latter case, when the elastic interaction between particles is an important factor (for example in metals), it is usually included as a bias in the capture distance between different particles.

Once all the transition rates $r_{ij}$ for all the possible states in the system are known (that is, they are given as input parameters) the OKMC algorithm starts. For simplicity we will omit the initial state in the transition rates and write them as $r_j$, being $j$ the final state achievable from a particular initial state $i$. Using this notation, the KMC direct method~\cite{GILLESPIE-JCP76} is applied as follows:

\begin{enumerate}
\item \label{ref:KMC1} Obtain the cumulative function 
\begin{equation}
R_i = \sum_{j=1}^i r_j
\end{equation}
for $i=1, \cdots, N.$ Being $N$ the total number of transitions in the given system.
\item Compute two random numbers, $r$ and $s$ in the interval $(0,1]$.
\item \label{ref:KMC3} Find $i$, the event to perform, for which $R_{i-1} < rR_N \le R_i$.
\item Perform the event $i$: transform the particular chosen object from $i$ to $j$.
\item Increase the total simulated time by
\begin{equation}
\Delta t = \frac{\ln(1/s)}{R_N}.
\end{equation}
\item Recalculate the affected rates.
\item Return to step \ref{ref:KMC1} until the requested physical time has been simulated.
\end{enumerate}

The above standard OKMC algorithm takes care of the time evolution only. Space dependence is intrinsic to the proper definition of each event. In our case the presence of physical defects that diffuse in space implies the need to include diffusion as a transition rate, and to define algorithms for space migration and particle interaction. Consequently, our OKMC simulator for damage evolution in solids contains the following modules:

\begin{itemize}
\item Objects (defects) and the list of their associated transition rates and actions.
\item A rate manager to compute time evolution and to pick up the event to perform.
\item A space manager to manipulate space translations, neighbor search and defect interactions.
\end{itemize}

\section{Implementation}
\label{sec:implementation}

\begin{figure}
\begin{center}
\includegraphics{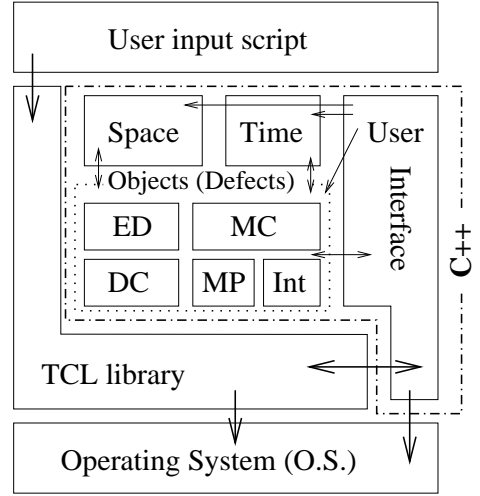}
\end{center}
\caption{Overall structure of the {\tt MMonCa} simulator. The user interface relies on a layer of the TCL interpreter,  extended to support OKMC. The extension relies on specialized modules to control the space, time and defects. Several defects are supported as the objects to be simulated: extended defects (ED), mobile particles (MP), damage clusters (DC), multi clusters (MC) and interfaces (Int).}
\label{fig:MMonCa}
\end{figure}

Fig.~\ref{fig:MMonCa} shows the overall structure of our simulator. {\tt MMonCa} has been implemented as C++ extensions of the TCL \cite{TCL} language. This allows us to use an already existing and well known language for the input script and to implement a user interface. The commands that have been extended allow the user to define a simulation cell, the 3D definition of the material structure of the simulation, reading of damage from an external file, annealing the damage, reading and writing the parameters needed for the simulation from the input file, and output of different quantities generated during the simulation, being the most important the concentrations and defect position and types. The rest of modules are described below.

\subsection{Computation of time: rate manager}
\label{sec:time}

\begin{figure}
\begin{center}
\includegraphics{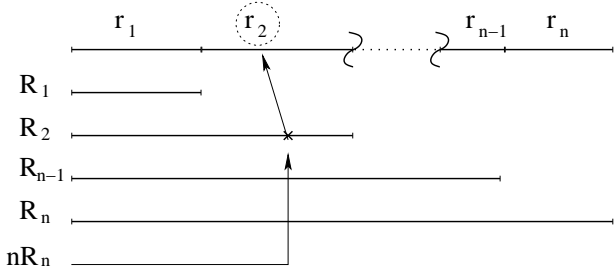}
\end{center}
\caption{The OKMC algorithm contains a list of all the transitions associated with the objects being simulated, and picks the next one proportionally to such rates. That can be seen graphically as getting a random number uniformly distributed in $[0,R_N)$ and picking up the event ``aligned'' with such number.}
\label{fig:scheduler}
\end{figure}

Fig.~\ref{fig:scheduler} shows graphically the idea behind the event selection involved in step \ref{ref:KMC3} of the OKMC algorithm previously explained. Once an updated list of all the transitions associated with the objects being simulated is generated, one of them is chosen proportionally to such rates. The $\Delta t=-\ln(s)/R_N$ associated with the simulation of such event is independent on the event chosen, depending only on the whole system. In practice, iterating through all the cumulative rates to find the one to be performed is not efficient when there is a large number of rates. For this reason, our simulator does not contain a transition bar with all the rates, but rather a binary tree, where the access time to each rate is not proportional to the number of them $N$ but to $\log_2(N)$. In the one hand, this improves the access time to the chosen event, on the other hand, the binary tree degrades the insertion, deletion and modification time for rate insertion from a constant time to also $\propto\log_2(N)$ time. Overall, the balance is positive when there are a large number of rates in the system and more selection of rates than insertions, modifications or deletions.

\subsection{Space organization and neighbor location}
\label{sec:space}

\begin{figure}
\begin{center}
\includegraphics{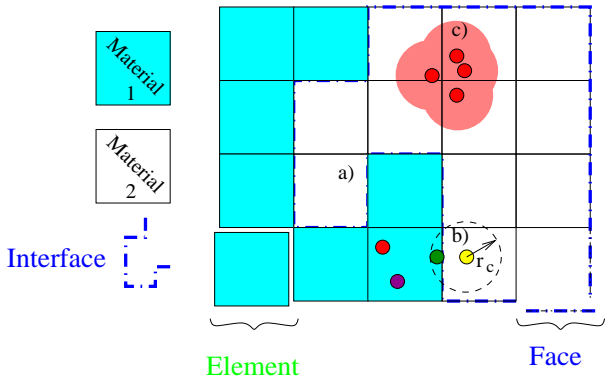}
\end{center}
\caption{Space is divided into small prismatic elements (rectangles in this 2D representation) using a tensor mesh. a) An interface is the union of all element faces between adjacent different materials. b) The capture distance $r_c$ of every particle is defined independently. c) The capture distance of clusters is built as the union of the capture distances of their constituent particles.}
\label{fig:space}
\end{figure}

The space is divided in small prismatic elements using a tensor mesh. Space is assumed to be homogeneous (material, temperature and other fields) inside each small element. Each element obtains its material definition by calling a user-defined procedure that allows the specification of the material structure in the simulation. This way, very complex shapes containing different materials can be simulated. When two consecutive mesh elements have different materials an interface object, as shown in Fig.~\ref{fig:space}a), is built between them.

Efficient neighbor search is implemented by having a standard link cell method~\cite{HOCKNEY-BOOK88}. Once the list of neighbors is obtained a look-up table is used to implement user-defined allowed interactions.

The capture distance $r_c$, shown in Fig.~\ref{fig:space}b), must be provided for every single particle. It is typically of the same order as $\lambda$, the microscopic migration distance. In our simulator, any non point defect (except interfaces) is created by the agglomeration and tracking of its constituent particles. This implies naturally that the capture distance of any defect is the overlay of all the capture distances of all the defect constituent particles as seen in Fig.~\ref{fig:space}c). It also means that extended defects can have any shape and a capture volume that will adapt to it as long as the particles are configured to form such shape.

\subsection{Defect structure}
\label{sec:defects}

We want to apply the theory of OKMC to the particular problem of simulating the evolution of damage introduced into a solid. Such damage can be introduced as an undesired side effect of the material application (for instance, when using the material in a fusion reactor) or it can be introduced on purpose to improve the material features (for instance, doping of semiconductors to produce devices). In any case, we define the objects of our simulator as the defects introduced in the material. In particular, we classify such defects as interfaces (Int), mobile particles (MP), damage clusters (DC), extended defects (ED) and multi-clusters (MC). The properties we simulate for each of them are described next. All these are considered objects of the OKMC simulator and are treated in a similar way. Each object, to be included in the simulator, needs to have the following data and functions defined:

\begin{itemize}
\item Number of events associated with the object. For instance, three for MPs: diffusion, breaking-up and creation of Frenkel pairs ($A \rightarrow A + I + V \rightarrow A_{i(V)} +V(I)$) to react with impurity atoms.
\item Rate associated to each event. In our example for MPs, computing the diffusion, breaking-up and injection rates, or returning zero if they do not apply.
\item Functions to perform each event when it is chosen by the OKMC algorithm. In the example, an MP needs the implementation to move the particle, break it up or create and/or react with Frenkel pairs.
\end{itemize}

Some of the explained events (break up and creation/reaction with Frenkel pairs) implement reactions similar to $AB \rightarrow A+B$ (for instance, $C_i \rightarrow C+I$ or $He_s \rightarrow He_i + V$). The forward reaction $A+B\rightarrow AB$ is implemented through diffusion. For this forward reaction to happen two things are needed: a) $A$ moving to the neighborhood of $B$, or $B$ into $A$, and b) the reaction being allowed. Diffusion is implemented as an event for all defects but interfaces. At the end of such event, a look for neighbors is performed to detect potential reacting species as explained in Sec.~\ref{sec:space}. To properly react with such species, two more algorithms are needed in each KMC object
\begin{itemize}
\item A look-up table that establishes whether the reaction is possible or not (taking into account possible reaction barriers)
\item A function that implements the interaction itself, taking the reactants and transforming them in the result.
\end{itemize}

Finally, since during reactions the reactants are destroyed and the result is created, each object requires a constructor and a destructor that is able to properly build and erase respectively the objects from the KMC simulator.

\subsubsection{Int: Interface}

\begin{figure}
\begin{center}
\includegraphics{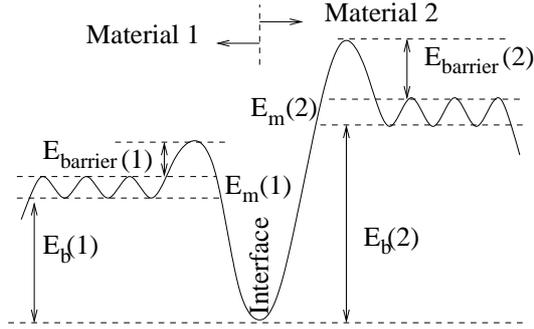}
\end{center}
\caption{Three phase segregation model. Particles can be captured and emitted at either side, but the binding energies, migration energies and capture barriers might be different at each side.}
\label{fig:interface}
\end{figure}

Fig.~\ref{fig:space} shows how the plane between two different materials or a material and the outside world is defined using an interfaces object. Interfaces can create and inject MPs (Is and Vs, and also impurities that were previously trapped). These emissions can be done to either side, assuming the MP may exist there. In the particular case of impurity emission, the model implemented corresponds to a three phase segregation model. Such model is shown in Fig.~\ref{fig:interface}. MP impurities can be at the interface by overcoming the barrier to reach the interface ($E_\mathrm{barrier} + E_m$). Then, they have a rate $\nu = \nu_0 \exp(-E_\mathrm{emit}(\mathrm{side})/k_BT)$ to be emitted to either side. $E_\mathrm{emit}$ is set as $E_b + E_\mathrm{barrier} + E_m$. Similarly to equation \ref{eq:equilibrium}, it is easy to see that the segregation coefficient, defined as the ratio between the concentration of particles at both sides at equilibrium, is 
\[S=\exp((E_b(1) - E_b(2))/k_BT).\]

When any diffusing defect arrives at the interface it can be annihilated according to certain probability set by the user. This applies to MPs, EDs, DCs and MCs.

\subsubsection{MP: Mobile particle}
Single ($He$, $I$, $V$) or paired defects ($CV$, $CI$) are defined as MP in our simulator, where paired interstitial defects are assumed to be the same as impurities in the interstitial position ($C_i = CI$). The transitions associated with these MP objects are:
\begin{enumerate}
\item Migration, by simulating the random walk of small diffusion events with fixed migration distance $\lambda$ in one of the three perpendicular axes of the system, randomly chosen for each jump. The migration rate for mobile particles is computed as $\nu = \nu_0\exp(-E_m/k_BT)$, where $\nu_0$ and $E_m$ are the input parameters for microscopic diffusivity. 
\item Break up of a pair (or kick off mechanism) of $I$ or $V$ impurities. For instance, $CV \rightarrow C + V$. The break-up frequency equals $\nu=\nu_0\exp(-E_{bk}/k_BT)$ with $E_{bk}$ being the activation energy for break-up. Such activation is computed as binding energy plus migration energy of the emitted particle.
\item Injection of extra Is or Vs by creating an IV pair, capturing the $I$ or the $V$ and emitting the other (also called Franck-Turnbull mechanism). This reaction applies for instance to $He$ substitutional in W: $He_s \rightarrow He_i + V$. Its rate is modeled as $\nu = \nu_0 \exp(-E_{FT}/k_BT)$ where the activation energy for such example would be set as $E_f(I) + E_f(V) - E_b(He_i) + E_m(I)$, being $E_f$ the formation energy.
\end{enumerate}

MPs can interact with each other to form more complex defect objects: for instance $I+I$ producing DCs or EDs, or $HeV + He$ producing MCs.

\subsubsection{DC: Damage cluster}

DCs are irregular agglomerations of $I$ and $V$ with a non-instantaneous recombination rate.  This mechanism simulates the recombination time needed by IV pairs in some systems, that although small, is not null, to annihilate both defects \cite{MOK-JAP05}. The rates associated with DCs are:

\begin{enumerate}
\item Recombination of IV pair with 
\[\nu = \nu_0\exp(-E_{IV}(\mathrm{size})/k_BT).\]
\item \label{DC1} Emission of MPs. The constituent particles can be emitted with a rate 
\[\nu = \nu_0 \exp(-E_\mathrm{emit}(\mathrm{size})/k_BT)\]
until the cluster dissolves. The activation energy for emission $E_\mathrm{emit}(\mathrm{size})$ is computed as the binding energy for each size plus the migration of the emitted particle.
\item \label{DC2} Transformation into an ED.  The transformation rate is computed as 
\[\nu_0 \exp(-E_\mathrm{transform}(\mathrm{size})/k_BT).\]
\item \label{DC3} Diffusion by random walk with rate 
\[\nu = \nu_0\exp(-E_m(\mathrm{size})/k_BT).\]
\end{enumerate}

Rates \ref{DC1}, \ref{DC2} and \ref{DC3} are non-null when the damage cluster contains only Is or Vs, but not both. DCs can also interact with MPs and EDs. 

\subsubsection{ED: Extended defect}

EDs are agglomeration of interstitials ($I_n$) or vacancies ($V_n$) with particular shapes that can emit their constituent particles, transform into other EDs, migrate and trap/detrap impurities that might stop their diffusion. In contrast with DCs, EDs contain only Is or Vs but never both. EDs can adopt different shapes to adapt to the realistic morphology of extended defects in different materials. In particular, they can be defined as a) planes (similar to \{311\} defects in Si \cite{MARTIN-BRAGADO-SSE08}), b) disks (similar to dislocation loops in Fe \cite{FU-NATURE05} or Si \cite{MARTIN-BRAGADO-SSE08}), c) spheres (voids in Si and other materials \cite{ESTREICHER-APL97}) and d) irregular clusters (no special shape). 

The transition rates defined for the different events are:

\begin{enumerate}
\item $\nu = \nu_0\exp(-E_\mathrm{emit}(\mathrm{size})/k_BT)$ for emission of MPs, being $E_\mathrm{emit}(\mathrm{size})$ the addition of binding energy plus migration energy of the emitted particle.
\item $\nu = \nu_0\exp(-E_\mathrm{transform}(\mathrm{size})/k_BT)$ for transformation into other EDs, being each transformation activation energy and prefactor defined by the user.
\item $\nu = \nu_0\exp(-E_m(\mathrm{size})/k_BT)$ for migration.
\item $\nu = \nu_0\exp(-E_\mathrm{detrap}(\mathrm{particle})/k_BT)$, one value for all sizes, to detrap previously captured particles. Independently the user can specify whether the trapped particles are only decorating the extended defects or also stops its diffusion. In this latter case, the detrapping rate plays a very important role for the overall diffusion rate of extended defects in the presence of traps.
\end{enumerate}

EDs can react with new incoming MPs. Depending on the nature of the incoming particle, the defects will grow ($I_n + I \rightarrow I_{n+1}$), annihilate it instantaneously ($I_n + V \rightarrow I_{n-1})$, or trap it ($I_n + C \rightarrow CI_n$). Reactions with other extended defects are also permitted ($I_n + I_m \rightarrow I_{n+m}$, $I_n + V_m \rightarrow V_{m-n}$ assuming $m>n$). They can also react with MCs (assuming that the final product is defined) and DCs and transform into a different ED with the same size (for instance, $I_n<111> \rightarrow I_n<100>$).

\subsubsection{MC: Multi-cluster}

MCs are the agglomeration of several impurities with either Is or Vs. They can play different roles in the physical systems under consideration. They allow the simulation of helium cluster formation in metals like Fe \cite{MORISHITA-NIMB03} and W \cite{BECQUART-JNM10}. In semiconductors, clusters of dopants with interstitials and vacancies deactivate partially the implanted dopants by forming agglomerations like $As_4V$ \cite{MUELLER-PRB03,PINACHO-APL05} or boron interstitial clusters \cite{PICHLER-BOOK04,PELAZ-APL97}.

The different events that MCs can perform are:

\begin{enumerate}
\item Emission of their constituent particles as MPs. The activation energy for emission is computed as the formation energy difference between the final and the initial state when positive, plus the migration energy of the emitted particle. The potential energies of all the clusters are required input parameters for the simulation.
\item Emission of constituent particles in pairs. For instance, $A_nV_m \rightarrow A_{n-1}V_{m-1} + AV$. The activation energy being $E_f(A_{n-1}V_{m-1}) + E_f(AV) - E_f(A_nV_m)$ when positive, plus $E_m(AV)$.
\item Injection of non-existing Is or Vs by Frenkel pair creation ($A_nI_m \rightarrow A_nI_{m+1} +V$ or $A_nV_m \rightarrow A_nV_{m+1}$).  The activation energy equals to $E_f(A_nI_{m+1}) + E_f(I) + E_f(V) - E_f(A_nI_m)$ when such value is positive (zero otherwise), plus $E_m(I)$ for injection of Is.
\item Migration. The migration rates are defined in a similar way to all the other migrations as 
\[\nu_0 \exp(-E_m(\mathrm{cluster})/k_BT).\]
\end{enumerate}

The number of different MP emission mechanisms for a simple MC can be high. For instance, $He_4V$ clusters can make transitions to $He_4 + V$, $He_3 + HeV$, $He_3V + He$, $He_4V_2 + I$ and $He_3V_2 + HeI$. For clusters breaking into elemental particles (MPs) the simulator also has to considered the migration energies of both constituents. For instance, two rates are needed for $He_2V \rightarrow He + HeV$, once considering the barrier of $E_m(He)$ and the other $E_m(HeV)$.

MCs can react with MPs, EDs and other MCs as long as the formation energy of the final result is included as a parameter. Even in those cases, a probability to reject the reaction 
\[P=\exp[(E_f^i - E_f^f)/k_BT]\]
is defined to account for the barriers involved in the formation of the new cluster. If $E_f^f < E_f^i$ the reaction always happens.

\section{Results}
\label{sec:results}

This section describes the validation of the code by comparing with theoretical values and experimental results or other simulations in three different materials: iron, silicon and tungsten.

\subsection{Theoretical results}
\label{sec:validation}

\begin{figure}
\begin{center}
\includegraphics{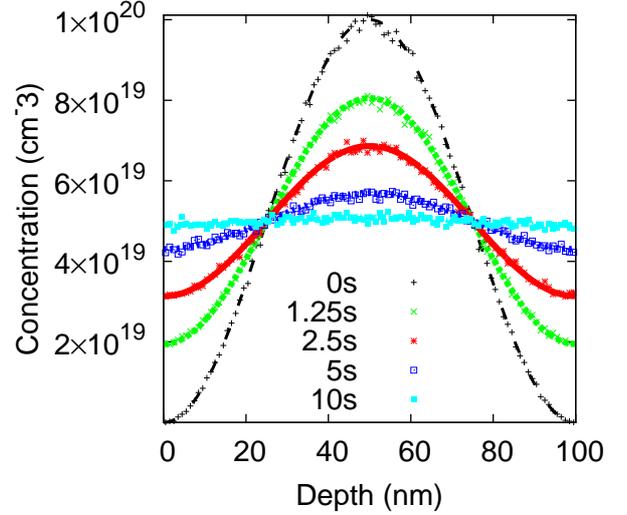}
\end{center}
\caption{(Color online). Comparison between KMC (symbols) and the theoretical solution (lines) for the time evolution of a spatial concentration of non-interacting particles with periodic boundary conditions.}
\label{fig:theory}
\end{figure}

Fig.~\ref{fig:theory} shows the comparison between KMC simulations (symbols) of the temporal evolution of an initial distribution of particles and the exact, theoretical results (lines). The calculation of the theoretical results has been done similarly to Ref.~\cite{MARTINEZ-JCP08}. The KMC simulations have been run for 10 seconds in a $100\times 300\times 300$\,nm$^3$ simulation cell with a total number of 448820 particles. The diffusivity of each particle was set to 100\,nm$^2$s$^{-1}$. Further comparisons with theoretical results, not shown here, have been done for interacting particles (for instance, diffusion of impurities through intermediate species $A + I \leftrightarrow A_i$ and break up), reaction with interfaces or sinks, correct establishing of equilibrium concentrations, etc.

\subsection{Iron}
\label{sec:Fe}

\begin{figure}
\begin{center}
\includegraphics{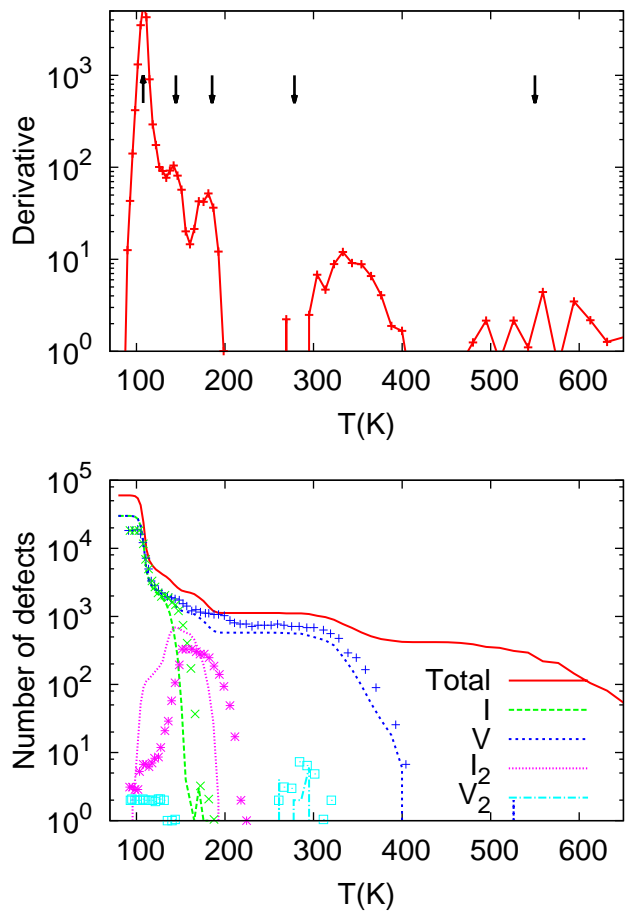}
\end{center}
\caption{(Color online). {\tt MMonCa} simulation of the evolution of defects and resistivity recovery during isochronal annealing. Top figure shows the derivative of the total defect concentration (red curve) being compared with experimental results~\cite{TAKAKI-RE83} for recovery stages (black arrows). Bottom figure shows the total simulated concentration of defects and the different defect contributions(lines) during the isochronal annealing of the sample after electron irradiation compared with previous theoretical work (symbols)~\cite{FU-NATURE05}.}
\label{fig:Fe}
\end{figure}

The study of defect kinetics in irradiated iron is a problem of primary importance for the aging of materials in the nuclear industry. Experimental work has been done by resistivity recovery experiments in high-purity electron-irradiated iron by Ref.~\cite{TAKAKI-RE83}, with irradiation doses in the range $\approx 2\times 10^{-6}$ to $\approx 200\times 10^{-6}$ displacements per atom (dpa). In these experiments, the resistivity of the metal is recorded during an isochronal annealing. The derivative of the resistivity versus the temperature shows clear peaks that are called recovery stages. These stages are related to different physical mechanisms involving the recombination, migration, growth and dissociation of the defects formed during irradiation and subsequent annealing. In particular, five important stages have been detected for iron.
\begin{itemize}
\item Stage I$_\mathrm{D2}$, observed at 107.5\,K related to the recombination of Frenkel pairs.
\item Stage I$_\mathrm{E}$ around 123 to 144\,K, as the result of the recombination of $I$ and $V$ belonging to different Frenkel pairs through the migration of interstitials. 
\item Stage II is suggested to happen when the $I_2$ starts to diffuse, around 164 to 185\,K. 
\item Stage III attributed to migration of Vs, around 220 to 278\,K.
\item Finally stage IV, around 520 to 550\,K produced by the dissociation of defect clusters formed during the previous stage III.
\end{itemize}

\begin{table}
\caption{OKMC Iron model}
\label{tab:Fe}
\begin{center}
\begin{tabular}{lcll}
\hline
Object & Migration & Species & Parameters \\
\hline
MB & Yes        & $I$ and $V$              & Ref.~\cite{FU-NATURE05,BJORKAS-PRB12} \\
ED & Yes        & $I_n$ small clusters     & Ref.~\cite{FU-NATURE05,BJORKAS-PRB12} \\
ED & Yes        & $<$111$>$ $I_n$ clusters & Ref.~\cite{BJORKAS-PRB12} \\
ED & size $< 5$ & $V_n$ clusters           & Ref.~\cite{FU-NATURE05,BJORKAS-PRB12}\\
\hline
\end{tabular}
\end{center}
\end{table}

Fig.~\ref{fig:Fe} shows the simulated isochronal annealing of $2\times 10^{-4}$\,dpa irradiated iron, together with the experimental stages (black arrows). It can be seen that the agreement with experiments \cite{TAKAKI-RE83} and with previous simulations done by other groups is good~\cite{FU-NATURE05,ORTIZ-PRB07}, especially taken into account that the compared results are produced by two different KMC methods (Event versus Object). A brief summary of the models and parameters used for such simulation is shown in Table.~\ref{tab:Fe}.

\subsection{Silicon}
\label{sec:Si}

The evolution of defects in silicon has been a subject of intense research for the past decades. Its interest relies on the need of semiconductor manufacturers to understand the Si system to produce more powerful electronic devices. One particular subject of study has been the characterization of damage by Si implantation. The evolution of such system contains many phases that are nowadays well known \cite{MARTIN-BRAGADO-SSE08}. The initial implantation produces a high population of Is and Vs, where the $V$ diffuses, even at room temperature implantations. During this initial stage Is and Vs do not recombine instantaneously, and tend to form DCs of various sizes. Depending on the particular implantation conditions, the amorphous pocket population might in some cases grow big enough to partially amorphize the sample. In other cases, dynamic annealing of the generated damage, that is, the annihilation of IV pairs during a cascade and the next one, might be enough to avoid amorphization. 

Once the implantation has finished, the sample is processed to anneal out the defects. This typically eliminates all the DCs, leaving only small extended defects in the beginning. Such extended defects are composed of the extra interstitials introduced by the implantation. During the annealing, the small, irregular interstitial clusters emit their constituent particles. This produces an almost conservative Ostwald ripening where big defects grow at the expense of small ones. At some point, the defects are big enough to be seen through the microscope, getting a characteristic \{311\} shape. Further annealing of these defects produces its dissolution or the formation of the very stable dislocation loops.

\begin{figure}
\begin{center}
\includegraphics{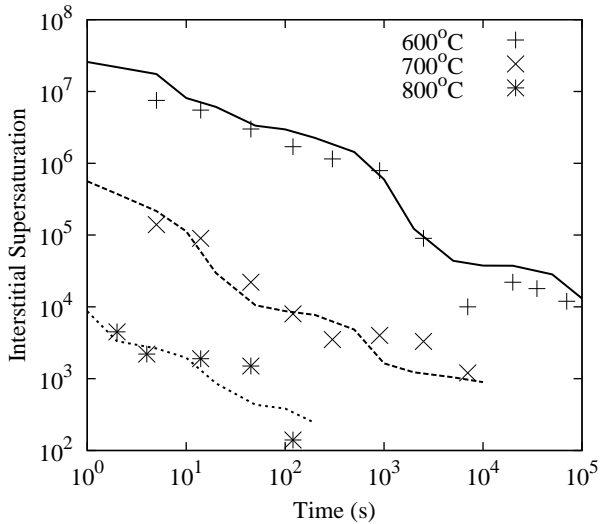}
\end{center}
\caption{Interstitial supersaturation as a function of time after a 40\,keV, $2\times 10^{13}$\,cm$^{-2}$ Si into Si irradiation at different temperatures 600, 700 and 800\,$^\circ$C. Symbols: Experimental data taken from Ref.~\cite{COWERN-PRL99}, lines: simulation results using the OKMC code {\tt MMonCa} presented in this work.}
\label{fig:supersaturation}
\end{figure}

Figure~\ref{fig:supersaturation} represents the comparison of experimental supersaturation (concentration of Is in equilibrium versus measured concentration) with the simulated results of {\tt MMonCa}. The experimental results are taken from Ref.~\cite{COWERN-PRL99}. In the experiment an implantation of 40\,keV, $2\times 10^{13}$\,cm$^{-2}$ Si$^+$ into Si was followed by annealing at 600, 700 and 800\,$^\circ$C. Table~\ref{tab:Si} shows the objects that we have defined and the references we use for the correct parametrization of such objects. Excellent agreement with both experimental data \cite{COWERN-PRL99} and simulations \cite{MARTIN-BRAGADO-SSE08} is achieved.

\begin{table}
\caption{OKMC Silicon model}
\label{tab:Si}
\begin{center}
\begin{tabular}{lcll}
\hline
Object & Migration & Species & Parameters \\
\hline
Int &     & $I$, $V$ creation & Ref.~\cite{SENTAURUS-12} \\
DC  & No  & $I_nV_n$          & Ref.~\cite{MOK-JAP05,SENTAURUS-12} \\
MP  & Yes & $I$ and $V$       & Ref.~\cite{BRACHT-PRL98, BRACHT-MRSB00} \\
ED  & No  & \{311\} $I_n$ clusters& Ref.~\cite{MARTIN-BRAGADO-SSE08} \\
ED  & No  & $V_n$ voids & Ref.~\cite{SENTAURUS-12} \\
\hline
\end{tabular}
\end{center}
\end{table}

\subsection{Tungsten}
\label{sec:W}

Tungsten is usually proposed as an appropriate material for nuclear fusion reactors due to a number of features: low-activation, high melting point, low sputtering yield, high thermal conductivity and low thermal expansion. W is proposed as armor material for inertial confinement fusion by laser with direct drive targets \cite{ALVAREZ-FED11}. For future magnetic fusion power plants W is considered the material of choice for the first wall and divertor \cite{BARABASH-JNM07}. Consequently, simulation of irradiation-induced damage in W by OKMC can help in the understanding of such a material \cite{RIVERA-NIMB13}.

The parametrization used to model W has been taken from Ref.~\cite{BECQUART-JNM10} and is summarized on Table~\ref{tab:W}. It constitutes a complex model that lets all defects interact with each other and with traps and allows for cluster formation. All pure clusters may migrate. In the particular case of interstitial clusters the migration is 1D along $<$111$>$ directions. Simulation boxes of dimensions $399\times 400\times 1001$ in lattice units, with lattice parameter $\lambda = 0.317$\,nm were used. The boundary conditions were periodic for $y$ and $z$. The x surfaces (both) were assumed to allow the desorption of incoming defects with a probability of 100\%: all approaching defects are annihilated.

\begin{table}
\caption{OKMC Tungsten model}
\label{tab:W}
\begin{center}
\begin{tabular}{lcll}
\hline
Object & Migration & Species & Parameters \\
\hline
MB & Yes  & $I$ and $V$    & Ref.~\cite{BECQUART-JNM10} \\
MB & Yes  & $He$           & Ref.~\cite{BECQUART-JNM10} \\
ED & Yes  & $I_n$          & Ref.~\cite{BECQUART-JNM10} \\
ED & Yes  & $V_n$          & Ref.~\cite{BECQUART-JNM10} \\
MC & Yes  & $He_n$         & Ref.~\cite{BECQUART-JNM10} \\
MC & No   & $He_nV_m$      & Ref.~\cite{BECQUART-JNM10} \\
MC & No   & $He_nI_n$      & Ref.~\cite{BECQUART-JNM10} \\
MC & No   & $CV_n$ (traps) & Ref.~\cite{BECQUART-JNM10} \\
MC & No   & $CI_n$ (traps) & Ref.~\cite{BECQUART-JNM10} \\
\hline
\end{tabular}
\end{center}
\end{table}

We compare our results with those of Becquart and co-workers for the $amorphous$ case \cite{HOU-JNM10}, i. e., we ignored the crystal structure of W when calculating the Frenkel pairs created by every incoming ion. 100\,appm of C were introduced as static traps acting on interstitials and vacancies, as well as on their clusters. We used the same irradiation conditions as Becquart (3\,keV $He$ irradiation at 5 K and 16 $He$ per second up to a dose of 12 ppm). We realized during the validation of {\tt MMonCa} that the results strongly depend on the initial conditions ($He$, $V$ and $I$ distributions). Therefore, we used the same initial defect distributions as Becquart and co-workers obtained for $amorphous$ W \cite{HOU-JNM10}. After the implantation the temperature was decreased to 1 K and isochronal annealing steps of 2 K for 60 s were simulated.

\begin{figure}
\begin{center}
\includegraphics{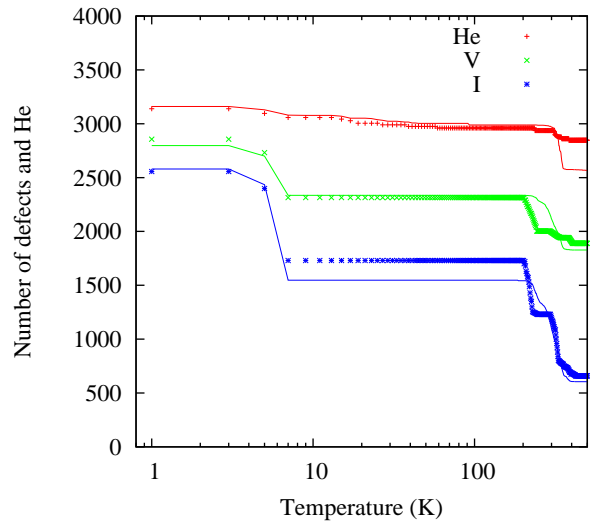}
\end{center}
\caption{(Color online). Comparison of the number of interstitial, vacancy and helium as simulated in this work (lines) and in Ref.~\cite{HOU-JNM10} (symbols).}
\label{fig:W-polycrystal}
\end{figure}

\begin{figure}
\begin{center}
\includegraphics{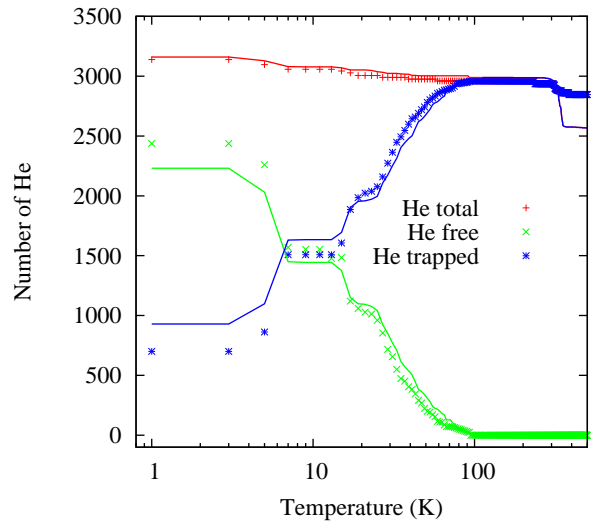}
\end{center}
\caption{(Color online). Comparison of the number of total, free, and trapped $He$ atoms as simulated in this work (lines) and in Ref.~\cite{HOU-JNM10} (symbols).}
\label{fig:W-amorphous}
\end{figure}

Fig.~\ref{fig:W-polycrystal} compares the results of {\tt MMonCa} (lines) with those presented in Ref.~\cite{HOU-JNM10} (symbols) concerning the evolution of interstitials, vacancies and helium remaining in the simulation box. Fig.~\ref{fig:W-amorphous}, on the other hand, displays the number of trapped and free helium atoms remaining after every annealing step. We can observe that the agreement is fair over the whole simulation for the different types of defects. Also defect clustering (not shown) is fairly reproduced. However, some discrepancies appear. We mainly attribute them to the different procedures employed by the codes to account for defect trapping. The code used by Becquart and co-workers considers that: (i) every defect has  an associated capture distance; (ii) clusters are spherical objects with an associated capture distance that in general grows with the number of constituents; and (iii) whenever the capture volumes (defined by the capture distance) of two defects overlap, trapping occurs. On the other hand, {\tt MMonCa} associates a capture distance to every single defect, whereas the clusters are formed by the agglomeration of single defects in different configurations (see Section~\ref{sec:space}). In any case, the identity of the single defects is not lost and trapping occurs when an object falls within a distance smaller than the capture distance of any single defect. Therefore, the trapping procedures are different and this turns out to be the source of the small discrepancies found when comparing the results. Note that the capture distances used in Table 5 of Ref.~\cite{BECQUART-JNM10} can not be directly used in {\tt MMonCa} because trapping is defined in different ways. In principle we must use capture distances approximately twice the size than those previously reported by Becquart and co-workers to account for their trapping criterion. We have found that the best results are obtained when we multiply the capture distances given in Table 5 of Ref.~\cite{BECQUART-JNM10} by 2.3 for the mobile particles $I$, $V$, $HeI$ and $HeV$ and by 1.5 for $He$, $C$, $CI$ and $CV$. With these values, {\tt MMonCa} slightly overestimates the interstitial loss at 7 K and the helium release at around 300 K Fig.~(\ref{fig:W-polycrystal}). In addition, the helium trapped fraction at low temperatures (Fig.~\ref{fig:W-amorphous}) turns out overestimated (the helium free fraction is underestimated). The different trapping procedures used in both codes are responsible for slightly different cluster formation during implantation. This, in turn, has consequences for the final evolution of the defects during the isochronal annealing. The largest differences between the codes are related to the evolution of big clusters, because the optimization of the capture distances can not account for the values assigned to every cluster by Becquart. However, despite the small discrepancies observed, we conclude that {\tt MMonCa} is able to reproduce complex results according to the expectations.

\section{Conclusions}
\label{sec:conclusions}

In this work we have reviewed the simulation techniques of the evolution of damage in irradiated solids and we have introduced the OKMC simulator {\tt MMonCa} and applied it to show the defect evolution in three different materials. We have started by explaining the theory of KMC and showing some details of how such theory has been implemented by creating generic structures and algorithms in the objects that we want to simulate. We have then reproduced experimental and simulated results in iron, silicon and tungsten using our simulator. The different comparisons show that {\tt MMonCa} can be successfully used to study the damage evolution of defects in solid materials validating the OKMC approach and the particular implementation into the {\tt MMonCa} simulator, that we hope will be of help for the materials research scientific community.

A copy of the simulator described in this work can be obtained at the following web page:

\url{http://www.materials.imdea.org/MMonCa/}

\section{Acknowledgments}

The authors thank the reviewers for their useful suggestions and insights. I. M.-B. wants to acknowledge funding of the project MASTIC (PCIG09-GA-2011-293783) by the Marie Curie Actions Grant FP7-PEOPLE-2011-CIG program.  This work contributes to the Joint Programme of Nuclear Materials of the European Energy Research Alliance. Thanks also to M. Jara\'iz and M. D. Johnson for the several fruitful discussions on KMC simulators during the past years.

%\bibliographystyle{model1a-num-names}
%\bibliography{../articles}

\end{document}